\documentstyle[floats,aps,aipbook,psfig]{revtex}
\righthead{J. P. Norris}
\lefthead{Canonical Timescales}
\begin{document}
\title{Canonical Timescales In GRBs--1995}
\author{Jay P. Norris}
\address{NASA/Goddard Space Flight Center, Greenbelt, MD 20771}

\maketitle

\begin{abstract}
Understanding the bimodal duration distribution (dynamic range $>$ 
10$^{4}$) of $\gamma$-ray bursts is central to determining if the 
phenomenon is in fact a singular one.  A unifying concept, beyond isotropy 
and inhomogeneity of the two groups separately, is that bursts consist of 
pulses, organized in time and energy:  wider pulses are more asymmetric, 
their centroids are shifted to later times at lower energies, and shorter, more 
symmetric pulses tend to be spectrally harder.  Long bursts tend to have 
many pulses while short bursts usually have few, relatively narrow pulses.
Two factors, viewing angle and beaming, may account for pulse asymmetry and 
the large dynamic range ($\sim$ 200) in pulse widths.  

A cosmological time-dilation signature, with an expected dynamic range of order
two, would be difficult to measure against these large intrinsic variations and 
low signal-to-noise levels of dimmer bursts.  Some statistics ($T_{\rm 90}$, 
pulse intervals) are particularly sensitive to brightness bias, noise, and 
apparently minor variations in definition.  Also, spectral redshift would 
move narrower, high-energy emission from dim bursts into the band of 
observation, constituting a countering effect to time dilation.  With analyses 
restricted to bursts longer than $\sim$ 2 s, tests for time dilation that are 
constructed to be free of brightness bias have yielded time-dilation factors 
$\sim$ 2--3, for pulse structures, intervals between structures, and durations.
\end{abstract}

\section*{GRB BIMODAL DURATION DISTRIBUTION}

Beyond the isotropy and departure from homogeneity of $\gamma$-ray bursts 
(GRBs), their bimodal duration distribution \cite{CK1} may be the most defining 
feature.  If we only understood why this phenomenon (or phenomena) has a 
dynamic range of almost five decades in event duration, we might begin to 
understand GRBs.  We do know a few things in connection with the bimodal 
appearance, enough perhaps to speculate that it reflects a unified phenomenon:  
Long and short bursts, on either side of the ``valley'' at $\sim$ 2 s, are 
separately isotropically distributed \cite{Briggs}.  There is evidence that 
both groups are undernumerous at low peak fluxes, compared to the Euclidean 
expectation in a homogeneously filled space \cite{Meegan}, although the
best measurement of peak flux for very short bursts is probably yet to be
realized.  That is, if one truly believes peak flux to be an indicator of 
distance with some fidelity, then time-tagged event data (2-$\mu$s resolution) 
should be employed to measure the peak fluxes of short bursts, since pulses 
in short bursts can be considerably shorter than the shortest (64-ms) 
timescale on which peak fluxes are tabulated in the BATSE 3B catalog.

Other indicators of kinship are found when the time profiles are examined in 
detail.  Both short and long bursts consist of pulses, that are organized in 
time and energy, as discussed in the ``Pulse Paradigm'' section below 
\cite{{JPN4},{JPN5}}.  In terms of spectral softening and asymmetry, pulses 
in short bursts appear to be carbon copies of pulses in long bursts, except 
that they are, on average, compressed by a factor of $\sim$ 20.  A possible 
difference between the two groups is that long bursts tend to have many pulses, 
whereas short bursts often have just a few major pulses structures \cite{JPN2}.
If the GRB phenomenon is singular, then ``telescoping'' distributions -- 
in pulse width, pulse interval, and number of pulses per burst -- 
may explain the valley and bimodal appearance \cite{{JPN2},{Wang}}.
Short bursts do tend to be spectrally harder \cite{CK2}.  But in general, 
bursts tend to soften as they progress, as demonstrated convincingly 
by Ford et al. \cite{Ford}.  Thus, the softer event-averaged spectra of long
bursts may be a matter of the radiation transfer of later pulses being affected 
by prior burst history.

\begin{figure}
\leavevmode
\psfig{file=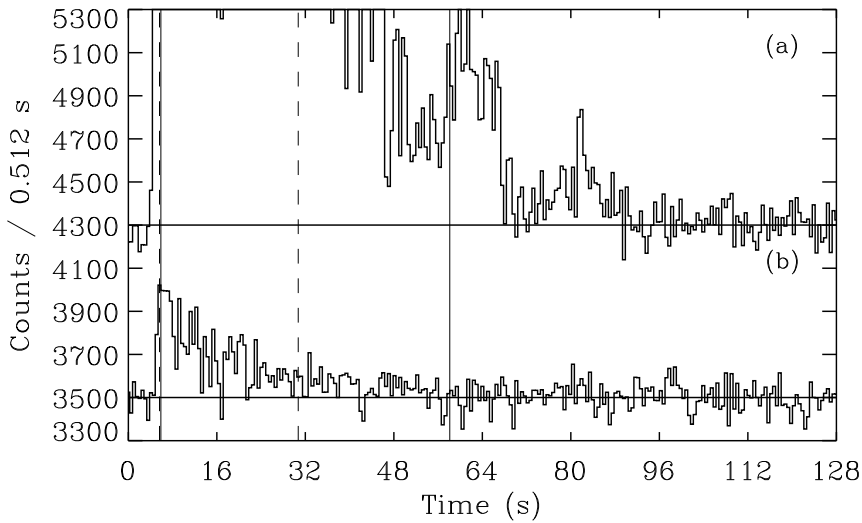,height=3.0in,width=5.0in}
\caption{
BATSE burst \# 678 with (a) original peak intensity, and with (b) peak 
intensity reduced to 1400 counts s$^{-1}$ and variance from the background 
interval of a dim burst added.  The $t_{\rm 5}$ and $t_{\rm 95}$ points are
shown as solid (a) and dashed (b) lines, determined from 4-$\sigma$ threshold
above background on timescales up to 16 s.}
\end{figure}

\begin{figure}
\leavevmode
\psfig{file=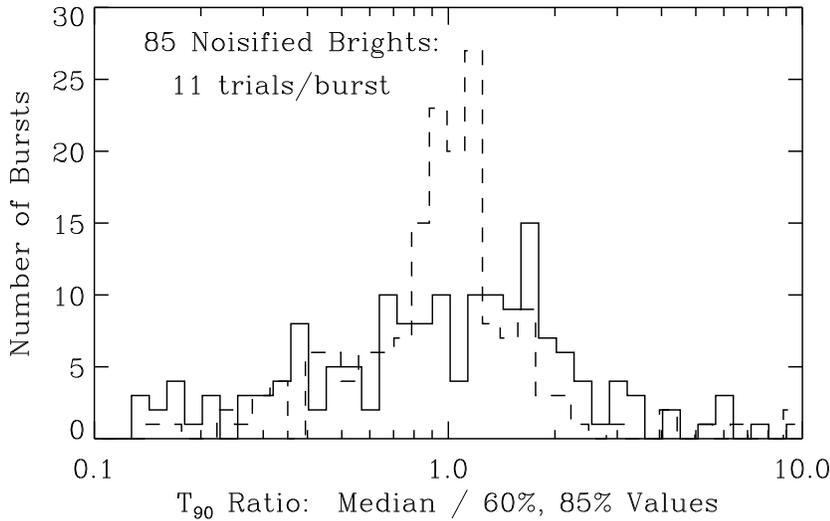,height=3.0in,width=5.0in}
\caption{
$T_{\rm 90}$ estimations for 85 long, bright bursts, 11 realizations per burst, 
with peak intensity and signal-to-noise equalized to dim burst level.  Solid 
and dashed histograms are median value divided by 1$^{\rm st}$ and 11$^{\rm th}$,
or by 3$^{\rm rd}$ and 9$^{\rm th}$ ranked values, respectively.  Note logarithmic
scale on abscissa.}
\end{figure}

Establishing duration measurements in a brightness-independent manner is 
desirable.  As illustrated in Figure 1, one can either have accurate durations 
for bright bursts, or estimations for all bursts (to some threshold) which are 
relatively free of brightness bias.  Figure 1 depicts BATSE trigger \# 678 
with (a) original peak intensity, and with (b) peak intensity reduced to 
1400 counts s$^{-1}$ and variance from the background interval of a dim burst 
added.  The $t_{\rm 0}$ and $t_{\rm 100}$ points are estimated by seeking the 
first and last fluctuation, respectively, to exceed the 4-$\sigma$ level above
background.  For the original peak intensity profile, due to the structure 
near bin $T$ = 60 s, $T_{\rm 90}$ is almost twice as long as that for the 
dimmed and noise-equalized profile.  Since for the 3B durations, $t_{\rm 0}$ 
and $t_{\rm 100}$ points were determined by eye and at original peak intensity 
and signal-to-noise (s/n) levels, one may conclude that durations of bright 
bursts are fairly accurately estimated, but that those of dim bursts are 
underestimated.  Since backgrounds are necessarily fitted to the original time 
profiles, there will be a tendency, regardless of succeeding steps in a 
duration measurement procedure, to declare low-intensity portions of the burst 
to be background intervals.  This problem exacerbates the underestimation of 
dim burst durations, and is particularly difficult to circumvent since it is 
desirable to specify the background near the burst, thereby more accurately
fitting any curvature.

The $T_{\rm 90}$ statistic is particularly vulnerable to statistical 
fluctuations, since the $t_{\rm 5}$ and $t_{\rm 95}$ points are necessarily 
near the $t_{\rm 0}$ and $t_{\rm 100}$ points, which are ill-defined.  Figure 
2 illustrates $T_{\rm 90}$ estimations for 85 long ($T_{\rm 90} >$ 2 s) bright 
bursts (peak flux [256 ms] $>$ 4.6 photons cm$^{-2}$ s$^{-1}$).  Eleven 
realizations were performed per burst, with peak intensity and s/n equalized 
to dim burst levels as described for Figure 1.  Solid and dashed histograms 
are for the median value divided by 1$^{\rm st}$ and 11$^{\rm th}$, or by
3$^{\rm rd}$ and 9$^{\rm th}$ ranked values ($\sim$ 85\% and $\sim$ 60\% 
confidence, respectively).  The spread in $T_{\rm 90}$ 
(note logarithmic scale on abscissa) is attributable in many 
bursts to low-intensity tails or small outlier structures, which are 
(de)accentuated by fluctuations present in one run but not in another, and 
which contribute little to the total counts, but considerably to the total 
event duration.  One must conclude that $T_{\rm 90}$ is not a terribly robust 
and useful statistic.  Obviously, $T_{\rm 50}$ is influenced by the 
determination of the $t_{\rm 0}$ and $t_{\rm 100}$ points, but less 
dramatically so.  Notice that minor variations in $t_{\rm 0}$ and $t_{\rm 100}$ 
will have a much larger relative effect on duration estimates for short bursts.

\section*{TIME DILATION}

In the early 1980s, upon discerning the quasi-isotropy of the KONUS 11/12 burst
localizations, Upendra Desai suggested to me to search for the signature of 
cosmological time dilation in the KONUS sample (duration distributions 
were constructed and nothing interesting was found).  In the BATSE era, 
Paczynski \cite{Pac} and Piran \cite{Piran} made quantitative predictions 
concerning time dilation in GRB time profiles, based on the BATSE isotropy 
picture, integral number-intensity relation, standard cosmology and simple 
source population assumptions (nonevolving luminosity [$L_{\rm GRB}$] and space 
density [$\eta_{\rm GRB}$] functions).  Because these assumptions have 
practically never been found to obtain for other cosmological populations 
\cite{{VT},{Hazard}}, we should not be surprised to find in the end that 
$L_{\rm GRB}$ and/or $\eta_{\rm GRB}$ evolve with cosmic time 
\cite{{Horack},{HMK}} or that any presently observed time stretching effects 
anti-correlated with GRB brightness are partly or purely special relativistic 
manifestations \cite{Brainerd}.

The timescales on which cosmological time dilation would be observed in 
GRBs and corresponding methods used to search for the signature are 
summarized in Table \ref{table1}.  Each timescale and method has its peculiar 
drawbacks, related to brightness bias; to the necessity of determining and 
applying a correction for spectral redshift of temporal structure; or to the 
heterogeneous nature of burst profiles.  It is difficult designing sensitive 
tests to measure a cosmological time-dilation factor, TDF = 
[1+$z_{\rm dim}$]/[1+$z_{\rm brt}$], which may be of order 
2--3 \cite{{Pac},{Piran},{JPN6},{JTB}}, when the dynamic range in durations 
is $>$ 10$^{4}$, and the dynamic range of pulse-structure widths is 
$\sim$ 10$^{2}$.

Assuming that the intrinsic burst process does not evolve with cosmic time, 
then all timescales in the table, except one, would necessarily be required to 
manifest equal TDFs after correction for redshift effects.  The exception 
(since $\eta_{\rm GRB}$ is not guaranteed a constant value) is the timescale 
between bursts, probed by the number-intensity relation.  Several studies 
\cite{{JPN1},{JPN3},{JPN6},{JPN7},{JTB}}, which take into account 
brightness-bias effects and energy-dependent narrowing effects arising from 
redshift have found mutually consistent results for long bursts, thus 
supporting the cosmological time-dilation interpretation.  For short bursts, 
several difficulties -- smaller sample with requisite temporal resolution, 
greater dispersion in relative duration from noise, difficulty of precise 
measurement of peak flux for pulses shorter than 64 ms -- combine to make 
bias-free analyses more problematic.  Recent automated approaches attempt 
to address these complications \cite{JDS}.

\begin{table}
\caption{Timescales for Cosmic Time Dilation in Long Bursts}
\label{table1}
\begin{tabular}{llll}
\multicolumn{1}{c}{Phenomenon} & {Timescale} & {Method}
 & {Primary Complications}\\
\tableline
Redshift & $\sim$ 300 keV/c 
 & peak in $\nu$F($\nu$) \tablenote{ref: \cite{Malozzi}} 
 & spectral shape, bandpass\\

Pulse widths & $\sim$ 0.2 -- 1 s & pulse fitting & ID'ing true pulses\\

Pulse Intervals & $\sim$ 0.3 -- 3 s & peak finding & ID'ing true intervals\\

``Burst Core'' & $\sim$ 2 -- 4 s & ACF & width correction\\
 & $\sim$ 8 -- 16 s & peak align & ID'ing peak, width corr.\\

Durations & 2 -- 600 s & $T_{\rm 50}$, $T_{\rm 90}$ & brightness bias\\

Burst Intervals & $\sim$ 1 day & log(N)-log(P)
 & $\eta_{\rm GRB}$, $L_{\rm GRB}$\\
\end{tabular}
\end{table}

As implied by Figure 1, a primary problem with using event durations to 
measure time dilation is that virtually all of a bright BATSE burst is easily 
apprehended by eye, whereas for a relatively dim BATSE burst ($\sim$ 20 -- 100 
times less intense), low-intensity structures are lost in the statistical 
fluctuations.  By constructing two duration distributions for long, bright 
bursts, with original peak intensity and with peak intensity equalized to that 
of dim bursts, and then applying a uniform threshold to define $t_{\rm 0}$ 
and $t_{\rm 100}$ points, we can roughly estimate the degree of brightness bias 
that must be inherent in the 3B durations.  Figure 3 illustrates these two 
distributions for the $T_{\rm 90}$ measure.  The distribution with equalized 
peak intensity is shifted to lower values relative to that for original peak 
intensity by factors of 1.7 (Gaussian fits), 1.4 (K-S test), or 2.8 (average 
of ratio of log[duration]).  The corresponding factors for $T_{\rm 50}$ are 
1.3, 1.2, and 1.7, respectively.  These brightness-bias factors will tend to 
obscure a time-dilation effect of order two.  Therefore, careful consideration
should be given to nullifying brightness-bias effects when estimating
durations.

\begin{figure}
\leavevmode
\psfig{file=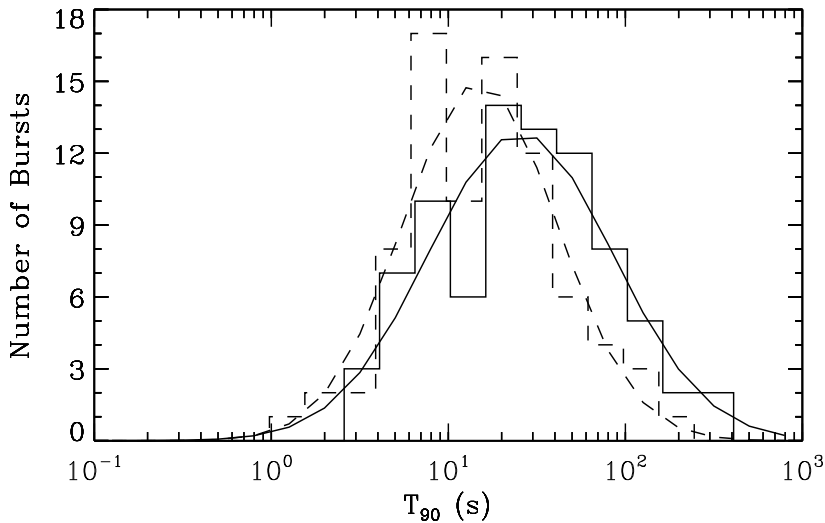,height=3.0in,width=5.0in}
\caption{
$T_{\rm 90}$ duration distributions and Gaussian fits for long, bright bursts, 
with original peak intensity (solid) and with peak intensity equalized to that 
of dim bursts (dashed).  The distribution for dimmed bursts is shifted to 
lower values, illustrating the brightness bias against ``seeing'' the 
entirety of dim bursts.}
\end{figure}

Two methods have been used to measure the width of the ``burst core'' --
the region near burst peak intensity -- often comprising several pulses.  
These are the peak-alignment method \cite{{JPN1},{Mitro}} and the 
auto-correlation function (ACF) \cite{{JPN6},{FandB}}.  
Such methods are widely appreciated to be common-sense approaches to 
search for time dilation in GRB profiles.  Both approaches utilize the region 
near maximum count rate.  The peak-alignment method is a linear sum of 
peak-normalized profiles, but finding ``the'' peak in dim bursts and 
accounting for noise bias at the peak can be problematic.  The ACF is a 
nonlinear combination of the peak region with itself.  The peak-finding 
problem is much ameliorated, but the ACF width is affected by noise in dim
bursts.  In both methods, the noise problems are addressable by equalization 
of the s/n levels of intensity groups that are being compared \cite{JPN6}.  

But, concomitant with time dilation is redshift:  the temporal structure 
associated with each energy range is shifted to lower energies in the 
observer's frame of reference.  Since GRB temporal structures are narrower 
at higher energy, this redshift-dependent narrowing of temporal structure 
competes with cosmological time dilation.  Moreover, in ``stack and 
average'' approaches, {\it dilation of intervals between pulses} is included, 
and this diminishes the original effect one wishes to measure, dilation of 
temporal structure.  Thus, width corrections must be applied in order to 
compare with expectations of any cosmological modelling.  The combined 
correction for redshift and interval dilation is largest for the ACF statistic, 
and practically as large for the peak-alignment method \cite{JPN6}.

Figure 4 is a schematic illustration of these two effects combined.  The top 
panel represents a burst at zero redshift.  The tallest pulse has been located 
to be used in some ``burst core'' measure of time dilation.  The bottom panel 
is the same burst coming from a redshift of unity, TDF = 2.  The solid curve 
shows the time-dilation effect.  Note that there is more space between pulses.  
The dashed curve includes the degree of pulse narrowing for a redshift of 
approximately unity.  Combined, interval dilation and pulse narrowing 
constitute a large width correction to the observed time-dilation measure.  
Dilation of intervals between pulses was not properly taken into account in 
previous treatments \cite{{JPN1},{FandB}}.  An estimation is made of the 
combined correction, using 16-channel data, by redshifting and time-dilating 
profiles of bright bursts, yielding correction factors of $\sim$ 1.4 
(peak-align) and $\sim$ 1.6 (ACF), assuming for example, an actual TDF of 
$\sim$ 2.5 between bright and dim bursts \cite{JPN6}.  Both approaches yield 
consistent corrected TDFs, but only constrain the allowable TDF to lie within
the range 2--3.

\begin{figure}
\leavevmode
\psfig{file=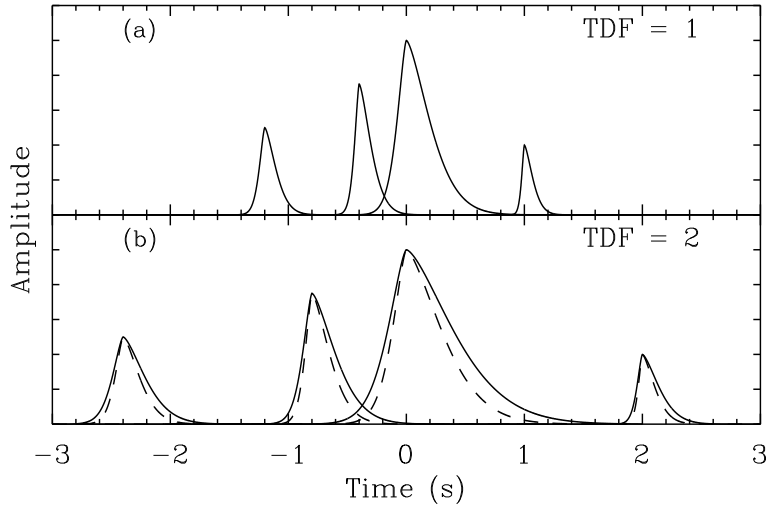,height=3.0in,width=5.0in}
\caption{
Schematic of dilation of pulses and intervals, and energy-dependent narrowing 
of pulse structure.  (a) Burst at zero redshift.  (b) Same burst from 
redshift of unity (TDF = 2); solid curve shows time-dilation effect on pulses 
and intervals; dashed curve includes pulse narrowing for redshift of $\sim$ 1.  
Combined, interval dilation and pulse narrowing necessitate large time-dilation
correction for methods which average time profiles or ACFs.}
\end{figure}

Note that since bursts are shorter at higher energy \cite{Ford}, a correction 
analogous to that required to compensate for pulse narrowing is necessary for 
duration measures of time dilation \cite{JTB}.  If the true TDF were 2 (or 3), 
then the correction factor is of order 1.10 (or 1.25).  Thus the time-dilation 
measure for durations is less affected by redshift than those for 
the ``burst core.''

For one potential measure of time dilation, [3B fluence]/[3B peak flux], 
three adjustments would be necessary to correct for weighting of the counts by 
the redshifted photon energy (in fluence but not flux), narrowing of pulse 
structure, and shorter durations at higher energy.  In fact, this measure would 
yield uncorrected TDFs {\it less than unity} for dim bursts.  In addition, 
systematic errors in fluence for dim bursts may arise in the process of 
background fitting.

One might think that measurements of intervals between pulse structures 
would be most free of systematics and energy-dependent effects, and thus 
would yield an observed TDF closest to the actual TDF.  But some 
consideration shows that related problems will plague interval measures of 
time dilation:  Intervals between pulses in bright bursts are much more 
clearly delineated than in dim bursts.  As a time profile is dilated,
additional, shorter intervals can appear near the limit 
of resolution of the data.  Also, narrower temporal structures at higher 
energy redshifted into the observation band will result in deeper minima 
between peaks, more readily sensed by a thresholding algorithm.  The latter
two effects will tend to result in a smaller observed TDF as two new, shorter
intervals arise from bifurcation of one interval.

Procedures have been devised in attempts to overcome these problems
\cite{{JPN7},{Neubauer}}.  Rendering all burst profiles to the same s/n 
level helps defeat the brightness-bias problem.  Wavelet- (or Fourier-) 
denoising of profiles \cite{Young} may be performed to diminish the 
frequency of insignificant peaks.  Approximate calibration of systematic 
effects arising from time dilation and redshift can be achieved by stretching 
and redshifting bright bursts at original resolution, and performing 
measurements on all bursts  at several times the original resolution.  Some 
attempts have been made to devise good automated tests of interval dilation, 
but these methods are still somewhat immature.  Distributions of pulse widths 
and intervals were measured by Davis using a semi-automated approach 
\cite{Davis}.  Scargle has developed a fully automated pulse-fitting algorithm 
that should allow model variations to be tested for short bursts \cite{JDS}.

If the time dilation observed in bursts is cosmological, then {\it intervals 
between bursts} would also be time-dilated, and this would be reflected in the 
number-intensity distribution since the observable is a ``density rate'' 
(e.g., bursts yr$^{-1}$ Gpc$^{-3}$) -- observed volumes and time intervals 
at high redshift are larger than in the comoving frame of the burst.  
These General Relativistic effects are correctly taken into account in some 
\cite{{Horack},{FandB},{CohenPiran},{RJN2}}, but not all, recent literature.  
However, the possibilities of cosmological source-density and/or luminosity 
evolution, or nonstandard cosmology, introduce considerable latitude into the 
modelling of redshift parameter space, enough so to make problematic -- or at 
least questionable -- the derivation of useful constraints vis-a-vis inclusion 
of time-dilation measurements.  Also, some treatments so far have 
(inconsistently) analyzed results of time-dilation measurements for long bursts 
($>$ 2 s) and the number-intensity distribution for all bursts.  Very 
interestingly, it may be that a GRB luminosity function is implied by the fact
that dim, soft-spectrum events appear to follow a number-intensity relation 
rather close to a -3/2 power law \cite{{Pizzi},{Belli},{CK1}}.  This could 
have important ramifications for many aspects of GRB analyses and modelling.

\section*{PULSE PARADIGM}

In the hard X-ray/low energy $\gamma$-ray portion of the spectrum -- now well-
mapped by BATSE -- GRBs appear to be composed of pulses which are organized in 
time and energy.  This observation is important because pulses must reflect 
the ``atoms'' of the emission process, the details upon which we must 
ultimately rely to understand much of the high-energy physics, even if bursts 
are eventually observed in other wavebands.  A global characteristic to be 
explained by physical modelling is that the envelopes of burst time profiles 
tend toward asymmetry (longer decays) \cite{Link}.  This tendency appears 
to extend to shorter timescales within a burst \cite{RJN1}, that is, to pulses.

Figure 15 of ref. \cite{JPN5} 
schematically illustrates the range of pulse shapes that are found in relative 
isolation, where we do not have to worry too much about blended, overlapping 
shapes distorting the statistics.  Shown are the approximate extremes of pulse 
shapes and spectral dependences of pulses that have been fitted in bright 
bursts longer than $\sim$ 2 s.  The range of pulse behavior is seen to go from 
\{symmetric, narrow, with negligible lag between energy bands, and tending 
to be spectrally harder\} to \{asymmetric with longer decays, wider, low 
energy lagging high energy, and spectrally softer\}.  Similar patterns are 
found in a preliminary study of pulses in short bursts using time-tagged event 
data, but with the pulse timescale telescoped by a factor of $\sim$ 20 compared 
to that for long bursts.  Pulses in short bursts are more often completely 
separated from their neighbors (frequently a short burst appears to comprise 
only a few pulses), and perhaps for this reason the pulse paradigm appears 
cleaner for them (see Figure 2 of ref. \cite{JPN4}).  Automated pulse fitting
and denoising approaches may clarify this picture \cite{{JDS},{Young},{Lee}}.  

Some bursts clearly exhibit the pulse paradigm throughout the whole event.  For 
example, trigger \# 2083 consists of two main pulses, both wide, with longer 
decays than rises, and with the pulse centroids shifted to later times at low
energy relative to high energy.  Trigger \# 678 (shown in Figure 1, but 
truncated and with insufficient resolution!) has a large number of spiky 
pulses, each one (as far as can be disentangled) symmetric down to a timescale 
of 16 ms with no shift in centroid with energy \cite{RJN1}.  Thorough analyses 
need to be conducted to determine to what degree this pattern prevails in 
bursts. 

The pulse asymmetry/centroid lag pattern extends across a range in pulse 
widths from $\sim$ 10 ms to $\sim$ 1 s.  It would seem that a variation in 
boost factor, $\Gamma$, cannot entirely account for the range of pulse widths, 
since then a comparable range in spectra would result.  But a combination of 
dynamic range in $\Gamma$ and variation in some geometrical factor, e.g., 
viewing angle, may account for the range of pulse widths.  In such a scenario, 
viewing angle would be related to degree of pulse asymmetry, which is 
correlated with pulse width in long bursts (for physical modelling of pulses, 
see refs. \cite{{FMN},{Mochko},{Shaviv},{Sari}}).  This would also be 
consistent with the observation that short bursts are, on average, spectrally 
harder than long bursts \cite{CK2}, and with the slight trend for more 
symmetric, narrower pulses to be harder \cite{JPN5}.  The observed spectral 
hardness of a pulse, however, must also be a function of intrinsic beaming 
factor and position within a burst (radiation transfer dependent on preceding 
pulses) since we know that bursts tend to soften as they progress \cite{Ford}.

In conclusion, I wish to point out that, once again at a GRB conference (last 
time, Stanford 1984), the temporal domain was given first billing.  Clearly 
this is more in tribute to BATSE's capabilities for recording bursts' 
time histories, allowing us to examine more closely their chaotic nature, 
rather than to our ability to understand the phenomenon.  But let us not forget 
that the burst physics is written in large proportion in their temporal 
evolution.

It is a pleasure to acknowledge innumerable enlightening conversations with
with many GRB aficionados, including especially Jerry Bonnell, Ed Fenimore, 
Robert Nemiroff, and Jeffrey Scargle.

\end{document}